\documentstyle[aps,prl,multicol,epsf]{revtex}
\def\be{\begin{equation}}
\def\ee{\end{equation}}
\def\ba{\begin{eqnarray}}  
\def\ea{\end{eqnarray}}

\def\bearst{\begin{eqnarray*}}
\def\eearst{\end{eqnarray*}}

\begin{document}
\twocolumn[\hsize\textwidth\columnwidth
\hsize\csname @twocolumnfalse\endcsname
\title
{
A cellular automaton model of gravitational clustering
}

\author
{
Roya Mohayaee and Luciano Pietronero
}
 
\address
{
Dipartimento di Fisica e unit\'a INFM, Edificio Fermi,
Universita Degli Studi di Roma "La Sapienza"\\
5, Piazzale Aldo Moro-100185, Roma, Italy
}

\maketitle
\begin{abstract}
Gravitational clustering of a random distribution of point masses is 
dominated by the effective short-range interactions due to large-scale isotropy.
We introduce a one-dimensional cellular automaton to reproduce this effect
in the most schematic way: at each time particles move towards 
their nearest neighbours with whom they coalesce on collision. 
This model shows an extremely rich phenomenology with
features of scale-invariant dynamics leading to a tree-like 
structure in space-time whose topological 
self-similarity are characterised with universal exponents. 
Our model suggests a simple
interpretation of the non-analytic 
hierarchical clustering and can reproduce some of the self-similar 
features of gravitational N-body simulations.

\end{abstract}

\hspace{.2in}
]


The basic mechanism of large-scale pattern formation 
by gravity remains mainly elusive, in spite of
elaborate studies in the past few decades \cite{saslaw99}.
Models of pattern formation, such as 
diffusion-limited aggregation (DLA) and its many variants
\cite{epv96}, which have been remarkably successful in 
wide ranges of disciplines,
do not appear to represent, even at a simplified 
level, gravitational clustering.
A different class of irreversible aggregation models, described by 
Smoluchowski coagulation equation
and its various associated kernels, gives rise to a power-law 
distribution of the masses of the 
aggregates which are, however, randomly distributed in space
\cite{smoluchowski16,ernst86,de86,sw78,sp97,aldous99}.

Gravitational clustering, in general, leads to a power law
distribution of the masses, known as the 
Press-Schechter mass function \cite{ps74}, but with a space distribution
which, at least up to some scale, is not 
homogeneous and can be described by fractal geometry
\cite{wlr99,cp92,smp98}. The key difference between  
the aggregations one encounters in statistical 
physics and gravitational clustering 
is that in the former, forces are 
short-ranged, and the nonlinear dynamics, driven by 
collisions, erase the memory of the initial conditions rather fast.
Gravity, on the other hand, is 
long-ranged with a deterministic evolution which 
strongly depends on the initial conditions 
even after the dynamics become nonlinear.

Despite these notorious features of gravity, it has long been shown by 
Chandrasekhar \cite{chandrasekhar43} that for 
the Poisson distribution of particle positions,
gravity is {\it effectively\/} short-ranged : it can be 
well-described by a simple nearest-neighbours interactions.
For the Poisson distribution of particles, the magnitude of 
the gravitational force, $\vert{\bf F}\vert$, has a 
Holtsmark distribution \cite{holtsmark17} given by
\ba
W(\vert{\bf F}\vert)
&=&
{2\vert {\bf F}\vert\over\pi}
\int_0^\infty
dy\, y\, {\rm sin} \left(y \vert {\bf F}\vert\right)
e^{-ay^{3/2}}
\nonumber\\
&\rightarrow & \sim{1\over \,\,\,\vert{\bf F}\vert^{5/2}}
\qquad\,\,{\rm as}\qquad\,\,  \vert {\bf F}\vert\rightarrow \infty
\label{holtsmark}
\ea
where $a=4n(2\pi G m)^{3/2}/15$, which is a constant depending 
on the mass of each particle, $m$, and the uniform 
number density $n$ \cite{chandrasekhar43}.
It was shown analytically and verified numerically that
the $5/2$ power-law tail of the Holtsmark 
distribution, is {\it exactly\/} due to the forces between 
the nearest neighbours \cite{chandrasekhar43}.
Even more remarkable is the fact that the distribution 
of the forces from the first neighbour approximation agrees 
with (\ref{holtsmark}) over most of the 
range of $\vert{\bf F}\vert$ \cite{chandrasekhar43}.
That the long-range gravitational 
force can be well-described by an effective 
short-range interaction may at first appear puzzling. However, the 
reason for this simplification is
that the isotropy of Poisson distribution is only 
broken at small scales where the granularity becomes important. 

As the system evolves, particle positions 
become correlated, but the self-similar nature of
gravitational clustering (see for example \cite{ps74,nfw95}) 
can ensure that at progressively larger scales
the functional form of the Holtsmark distribution remains intact.
Hence, under an appropriate renormalisation of the 
masses and the distances, determined by 
the scale of granularity at a given time, 
gravitational interactions can be taken to be
effectively short-ranged, even 
at times comparable to the dynamical time of the system.

Inspired by these facts, we introduce an extremely simple
one-dimensional automaton model which 
focuses precisely on this granular and self-similar nature 
of the gravitational clustering phenomenon. 
Unlike most previous models such as DLA, 
Smoluchowski or 1-dimensional Burgers, our aggregation 
rule is position-dependent and independent of 
the masses and initial velocities of the aggregates.
Unlike the conventional boolean cellular automata, we distribute particles
randomly on a line rather than on discrete lattice points.
We focus mainly on the universal features of the distribution of the masses of
the aggregates and on the space-time topological properties of the full 
aggregation history, {\it i.e.}, on the universal characteristics of 
the ``merger tree'' of our model. 

\begin{figure}[hbt]
\centerline{
        \vspace{-0.8cm}
        \epsfxsize=0.44\textwidth
        \epsfbox{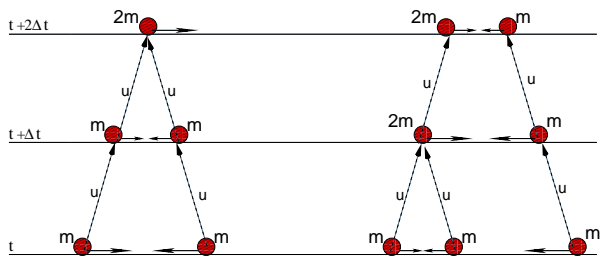}
           }
\vspace{1.0cm}
\caption
{
A schematic illustration of the aggregation rule (\ref{rule}). 
The essence of our model is that particles move towards their nearest
neighbours and coalesce on collision.
}
\label{schematic}
\end{figure}

The basic idea of this model, that the particles move
towards their nearest neighbours, is implemented by the following 
algorithm.
We distribute $10^5$ particles randomly on a periodic line
with a length of $10^5$ units, so that the average density is equal to one. 
Particles move towards their nearest neighbours either by one unit 
at each time step or by half their separations, whichever that is shorter. If
two particles are closer than a lower threshold, which we set 
equal to 2 units, and, in addition, are {\it mutual} nearest-neighbours they 
coalesce at the mid-point of their separation, conserving mass.  
This aggregation rule for a particle $i$ at the position $x_i(t)$ at time $t$ 
can be written as,
\ba
x_i(t+\Delta t)&=&x_i(t)+
{\rm Min}\left[{x_r(t)\over 2},1\right]
\quad\,\,\,\,
{\rm If}\,\,\,
x_r(t) < \vert x_l(t) \vert
\nonumber
\\
x_i(t+\Delta t)&=&x_i(t)+
{\rm Max}\left[{x_l(t)\over 2},-1\right]
\,\,\,\,
{\rm If}\,\,\,
x_r(t) > \vert x_l(t) \vert 
\nonumber\\
\label{rule}
\ea
where $x_r(t)=x_{i+1}(t)-x_i(t)$ and $x_l(t)=x_{i-1}(t)-x_i(t)$. This process
is illustrated in Fig.\ \ref{schematic}.

\begin{figure}[htb]
\centerline{
        \vspace{-0.8cm}
        \epsfxsize=0.44\textwidth
        \epsfbox{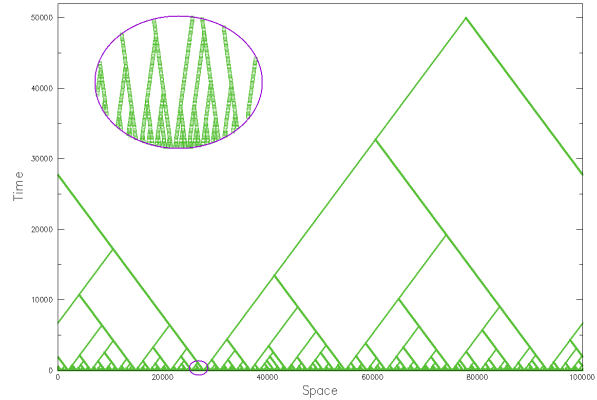}
           }
\vspace{1.0cm}
\caption
{The merger tree: the trajectories of $10^5$ particles from 
a run of our simulation. The space-time merger tree 
exhibits the property of {\it topological scaling}, a concept
which is relevant for branched structures and originally arose 
as a means of analysing river networks in two-dimensional space.}
\label{tree}
\end{figure}

As time progresses, masses and distances of the nearest neighbours 
rescale and the same rule, given by (\ref{rule}), is followed by more massive 
and farther apart aggregates.
The continuation of this process will eventually lead to 
the total coagulation of the colloidal 
particles into one single mass.
This aggregation mechanism traces a self-similar tree-like 
structure in space-time as shown in Fig.\ \ref{tree}.

The tree structure of the aggregation process in space-time 
is a manifestation of topological self-similarity \cite{hasely2000}, which is 
a property of many branched structures such as river-networks 
\cite{rinaldo97} and bronchial trees \cite{wbe97} and can be quantified by 
various scaling exponents. A most common of these exponents is the Strahler 
index, $s$, given by
\be
P(\eta)\sim \eta^{-s}\, ,
\label{strah}
\ee
where $P(\eta)$ is the probability that a point in the network 
is connected to $\eta$ other points 
uphill, also known as the drainage area. 
The Strahler index, $s$, is a measure of the 
bushiness of a branched structure and 
has an upper limit of $1.5$ for river networks \cite{rinaldo97}.

\begin{figure}[htb]
\centerline{
        \vspace{-0.8cm}
        \epsfxsize=0.44\textwidth
        \epsfbox{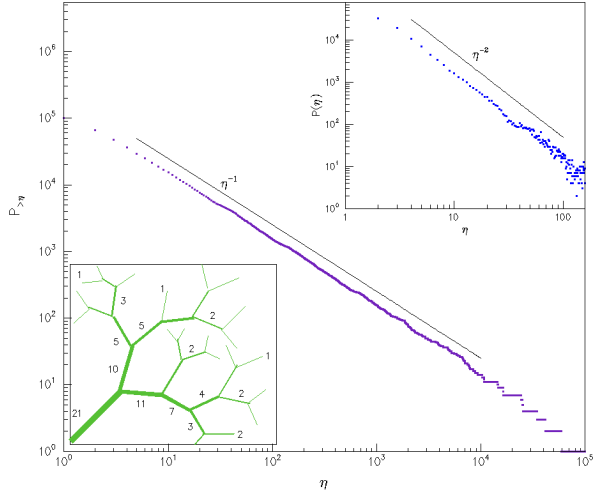}
           }
\vspace{1.0cm}
\caption{
Topological scaling: the scaling, over many decades, of the probability 
distribution of the masses of the newly-borned aggregates, $P(\eta)$,
(upper right inset) and of the corresponding cumulative probability,
$P_{>\eta}$,
{\it i.e.} the probability that the mass of a newly-formed aggregate is 
larger than $\eta$, (main plot).
The sketch in lower left inset illustrates 
the standard procedure of Horton-Strahler 
ordering and link amplitude to find the 
Strahler index: $P(\eta)$ is the frequency 
of occurrence of a number shown in that sketch.
}
\label{river}
\end{figure}

In our model, the function $P(\eta)$, given by (\ref{strah}), interprets as the 
probability that a newly-formed aggregate has a mass equal to $\eta$.
We observe a larger value for Strahler index
($s=2$, given by the slope of the plot in the 
upper inset of Fig.\ \ref{river}) than is 
expected for river networks and which seems to be 
a characteristic of Cayley tree structures. 
A notable example of such a 
structure, also with exponent $2$, has recently been observed 
for the property of internet connections\cite{cmp00}.   
The difference between our model and 
the conventional river-networks could be
the stringent requirement of mass conservation here. 

Topological scaling, is believed to emerge from a self-similar growth process.
Self-similar growth, or dynamical scaling, is a dominant feature of
hierarchical gravitational clustering \cite{nfw95},
and is the fundamental reason for the 
emergence of distribution functions such as 
Press-Schechter mass function in cosmology\cite{ps74}. 
An appropriate way to analyse dynamical scaling is, indeed, to study 
the mass distribution function
$n(m,t)$. Note that our function differs from the usual number 
density by a constant factor
which is given by the size of our system.

It can be easily inferred from Fig.\ \ref{tree}, and we have 
verified numerically that
the average mass, 
$\langle m\rangle$, and the mass variance, 
$(\langle m^2\rangle-\langle m\rangle^2)^{1/2}$ grow linearly with time. 
It is worth comparing this with the growth of average mass in one-dimensional 
Burgers, for example with uniform 
initial velocities and positions, where an exponent of $2/3$ 
has been obtained \cite{cpy90,bcdr99}.
The linear growth of average mass also holds for Smoluchowski equation with 
a constant kernel \cite{aldous99}. 
The universality between our model and Smoluchowski 
arises inspite of the fact that particle
trajectories are not Brownian here. 
In addition to this common feature, our model and Smoluchowski equation have 
similar asymptotic states. In both of these models, clusters collide until 
all the mass falls into one final clump, whereas in 1-dimensional 
Burgers, the asymptotic state can contain many clumps with zero momentum.

\begin{figure}[htb]
\centerline{
        \vspace{-0.8cm}
        \epsfxsize=0.44\textwidth
        \epsfbox{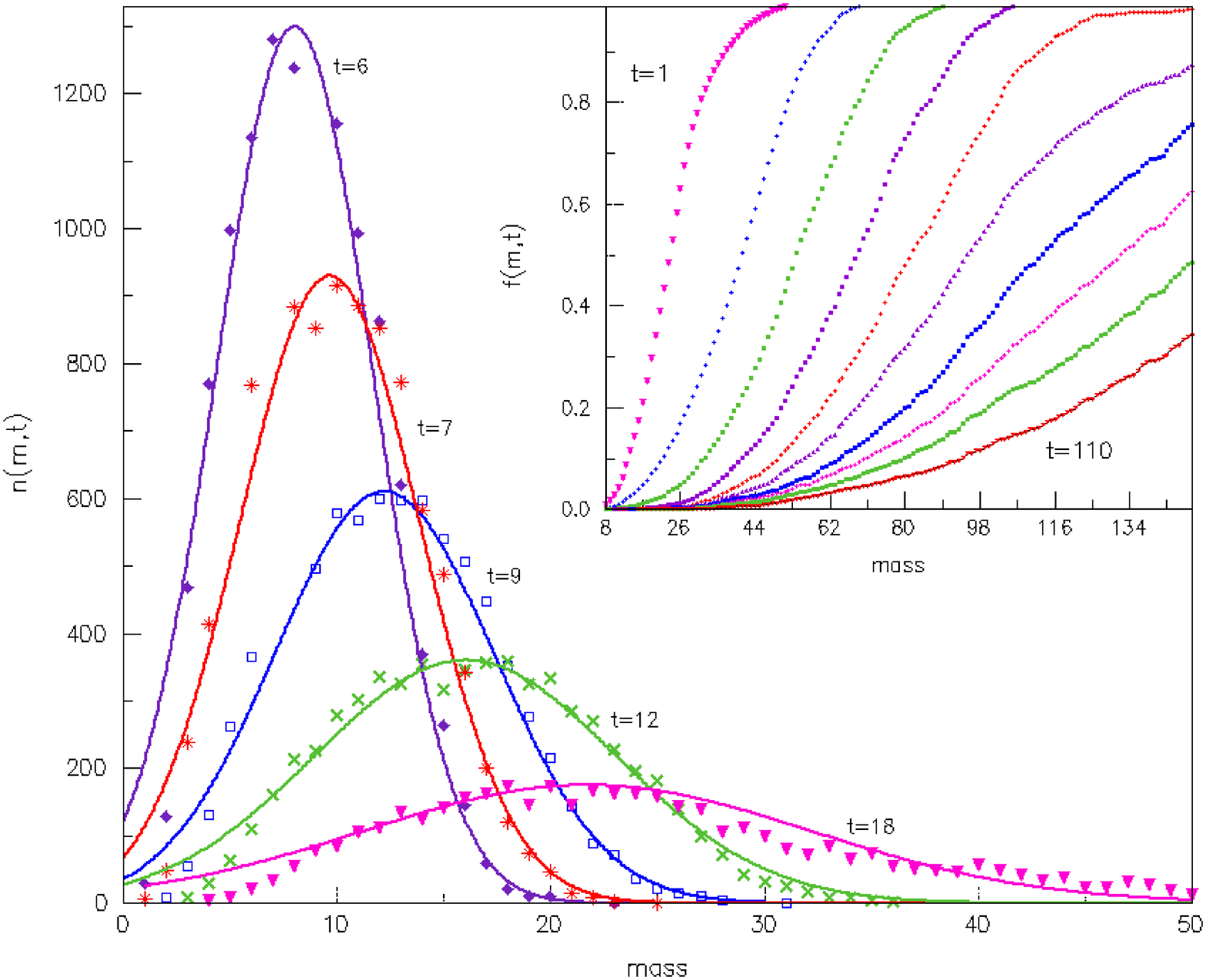}}
\vspace{1.0cm}
\caption{
Dynamical scaling: self-similar growth of mass distribution 
function $n(m,t)$. Fits on the main plot are obtained with
a simple Gaussian. The inset shows the self-similar 
growth of the fraction, $f(m,t)$, of mass
in objects of mass smaller than $m$, 
spanning over the significant part of the dynamical time.}
\label{selfsimilarmass}
\end{figure}

The distribution of the masses deviates slightly 
from a simple Gaussian for 
small masses where it develops a power-law
as is shown in Fig.\ \ref{selfsimilarmass}.
The inset clearly demonstrates that a self-similar condensation
process, rather similar to the one 
observed in gravitational 
N-body simulations (compare with Fig.\ 1 of \cite{ps74}),
has set in. In the process of coagulation, the shape of 
the mass spectrum seems to become fixed, and the curves move in parallel 
to the right (increasing aggregate mass). 

The results presented so far provide strong evidence 
that the mass distribution function has the general scaling form:
\be
n(m,t)\sim 
m^\alpha t^\beta \exp \left(-m^\gamma/t^\lambda \right)\,,
\label{scaling}
\ee
where the value of the exponents $\alpha,\beta,\gamma$ 
and $\lambda$ will be found in what follows. 

The conservation of the total mass in our model
leads to the exponent identity
\be
\beta+{\lambda\over\gamma}\left(2+\alpha\right)=0\, .
\label{identity1}
\ee

In addition, the linear rate of growth of the average mass,
$\langle m\rangle\sim t$, leads to the further scaling identity,
\be
{\lambda\over\gamma}=1\, .
\label{identity2}
\ee

At this point one can already see the emergence of 
a topological scaling, namely a power-law behaviour in the time-integrated
mass distribution function, at small masses. We comment 
that this is different from Strahler index we found
in Fig.\ \ref{river}, since the latter refers only to newly-formed aggregates.
This topological scaling, namely the power-law behaviour at small masses of
the time-integrated mass distribution function, 
$
\int n(m,t) dt\sim {1/ m^l}\, ,
$ 
is implied by the scaling relation (\ref{identity1}), 
which fixes the value of the exponent $l$ to $l=1$. 
This value is also confirmed by our numerical simulations. 

A further test of our scaling identities 
(\ref{identity1}) and (\ref{identity2}), is provided by
the evolution of the maxima, $n_{\rm max}$, of the mass 
distribution function, {\it i.e.}, by the rate of decay of the peaks of 
the main plot in Fig.\ \ref{selfsimilarmass}. Using the fact that 
the average mass grows linearly with time, in the scaling 
expression (\ref{scaling}), we obtain 
$
n_{\rm max}(m,t)\sim {1/ t^2},
$
which is again confirmed by our simulations.

Thus, the scaling identities 
(\ref{identity1}) and (\ref{identity2}), reduce our scaling ansatz 
(\ref{scaling}) to the self-similar form
\be
t^2 n(m,t)\sim
\left({m\over t}\right)^\alpha
e^{-(m/t)^\gamma}\,.
\label{iden}
\ee
The factor of $1/t^2$ in (\ref{iden}) is a consequence of mass conservation
and the linear growth rate of the average mass, 
and has also been observed for the constant-kernel solution to Smoluchowski 
equation \cite{aldous99}:
$n(m,t)=4 t^{-2}\exp(-2m/t)$. This solution of Smoluchowski 
equation, is obtained
from our general solution (\ref{iden}) 
by the transformation $t\rightarrow 2t$ and by using 
the following values of the exponents: 
$\alpha=0$, $\gamma=1$. We shall soon show 
that these exponents take different values in our model.

\begin{figure}[htb]
\centerline{
        \vspace{-0.8cm}
        \epsfxsize=0.44\textwidth
        \epsfbox{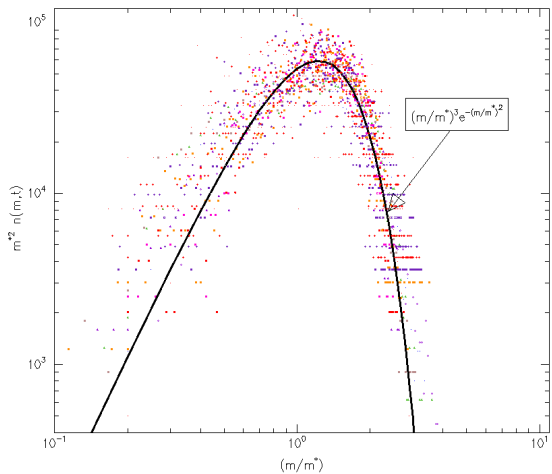}}
\vspace{1.0cm}
\caption
{
The mass function: the time-rescaled
mass distribution function, 
$n(m,t) {m^*}^2$, scatters around the fit (the continuous curve) 
for more than two decades in time.
The characteristic cutoff mass, $m^*$, grows linearly with time. 
}
\label{scalingfig}
\end{figure}

To put our notation in accordance with that 
used in cosmology, we replace the time variable 
by the cut-off mass $m^*(t)$. As we have mentioned 
previously, our simulations show that a
typical mass grows linearly 
with time, {\it i.e.} $m^*\sim t$.
The exponents $\alpha$ and $\gamma$ in expression (\ref{iden})
can be found by plotting $n(m,t){m^*}^2$ against the ratio $m/m^*$,
which we have done in Fig.\ \ref{scalingfig}.

We show in Fig.\ \ref{scalingfig} that for over 
two decades in time scale 
the functional form of the mass 
function, given by the RHS of (\ref{iden}), is preserved.
The fit in Fig.\ \ref{scalingfig} sets the value 
of the unknown exponents in (\ref{iden}) to
$\alpha\sim 3$ and $\gamma\sim 2$ and we finally arrive at
the approximate expression
\be 
n(m,t)\sim {1\over {m^*(t)}^2} 
\left({m\over m^*(t)}\right)^3 e^{-(m/m^*(t))^{2}}
\label{scalingfinal}
\ee
for the mass distribution function.
We emphasis that, our solution is different
from the constant-kernel solution to 
Smoluchowski equation which is an
special case of our general scaling 
solution (\ref{iden}), as previously noted. 
We comment that unlike some of the previous aggregation models
used in cosmology \cite{sw78,sp97}, our distribution function 
cannot be formed from a white noise initial spectrum. It remains
to be seen if, as for Smoluchowski with additive kernel, 
the introduction of a mass-dependent factor
in our aggregation rule (\ref{rule}), would give rise to a
Press-Schechter type mass function.

We have also analysed the density-density 
correlation function which 
has a power-law behaviour with exponent $-1$ at small 
scales, indicating 
that the mass is distributed on
zero-dimensional objects, and 
a crossover to a constant value at 
large scales where it
reminisces the initial Poisson 
distribution. The crossover length increases linearly
with time and the growth of the structures 
is dominated by the granular properties
which are shifted from small to large 
scales. Thus, our model differs from the usual 
statistical models which generate fractals, where
large-scale structures are built while 
substructures are preserved and 
not destroyed as is the case here.  
In this sense the present model 
does not generate asymptotic fractal structures in space.

In conclusion, the seminal result
of Chandrasekhar, that the long-range gravitational interactions
between randomly distributed particles can be 
almost exactly replaced by nearest-neighbour 
interactions, stimulated us to present a simple aggregation model
which captures this profound feature of gravitating systems.
We have shown that our model exhibits topological 
self-similarity over many decades of mass scale 
and dynamical scaling over many decades of temporal scale. 
These properties make it a simple
and intuitive model for the study of 
gravitational hierarchical clustering. 
  
We thank J.\ Bertoin, G.\ Caldarelli, U.\ Frisch, N.\ Menci, 
G.\ Murante, R.\ Sheth and A.\ Stella for 
useful discussions and suggestions. R.\ M.\ was supported by TMR 
network ``Fractal structures 
and Self-organization'' under the contract FMRXCT980183.


\end{document}